\documentstyle[12pt,aasms4,psfig]{article}
\begin{document}
\title{LARGE SCALE BIASING AND THE PRIMORDIAL GRAVITATIONAL POTENTIAL}
\author{Jounghun Lee and Sergei F. Shandarin} 
\affil{Department of Physics and Astronomy,\\
University of Kansas, Lawrence, KS 66045 \\
taiji, sergei@kusmos.phsx.ukans.edu}

\begin{abstract}

We modified the Press-Schechter (PS) formalism and then 
analytically derived 
a constrained mass distribution function $n(M|\varphi)$ for the regions
having some specified value of the 
primordial gravitational 
potential, $\varphi$.
The resulting modified PS theory 
predicts 
that gravitationally bound clumps with masses corresponding 
to rich clusters 
are significantly biased toward the regions of negative 
primordial potential - the troughs of the potential. 
The prediction is quantitative, depending on the mass and the depth 
of the troughs, which can be 
tested in large N-body simulations. 
As an illustration of the magnitude of the effect we 
calculate the constrained mass function for the CDM model
with $\Gamma = \Omega h = 0.25$ normalized to $\sigma_{8} = 1$.
In particular, we show that the probability
of finding a clump of mass  
$10^{14} - 10^{15}h^{-1}M_{\odot}$ in the region of negative initial
potential is $1.3 - 3$ times greater (depending on the mass) 
than that in the region of positive
initial potential. 
The scale of the potential fluctuations 
$R_{\varphi}=\sqrt{3} \sigma_{\varphi}/\sigma_{\varphi'}$ 
is shown to be
$\approx 120 h^{-1}{\rm Mpc}$ for the spectrum in question. 
The rms mass density contrast on this scale
is only about $\sigma _{\delta}(R_{\varphi}) \approx 0.03$. 
Assuming that the modified PS theory is statistically correct, 
we conclude that  
clusters are significantly biased 
($b \ge 10$, $b$ is a bias factor defined by 
$\Delta n_{cl}/ n_{cl} =b \Delta \rho_m/\rho_m$) 
toward the regions having negative initial 
potential. 

\keywords{cosmology: theory --- large-scale structure 
of universe}

\end{abstract}

\section{INTRODUCTION}
  
Assuming the standard hierarchical 
model of the structure formation from Gaussian
fluctuations due to gravitational instability, 
we study the effect of primordial gravitational potential 
fluctuations on massive objects such as galaxy clusters and perhaps 
superclusters, i.e., clusters of clusters (\cite{bah-son84}). 
We employ the PS formalism as a tool and 
modify it for this study.

Some effect of the primordial 
gravitational potential upon the structure formation  
has been already noted.
\cite{kof-sha88} have noticed that the adhesion approximation 
predicts that the formation of voids 
is associated with positive peaks of the primordial 
gravitational potential. Sahni, Sathyaprakash, \& Shandarin (1994)
studied the effect  and measured a significant correlation between 
the sizes of voids and  the value of primordial gravitational potential
in numerical simulations of the adhesion model.
By investigating the evolution of 
correlation between the potential and the density perturbations, 
Buryak, Demianski, $\&$ Doroshkevich (1992) showed that 
the formation of super large scale structures 
is mainly determined 
by the spatial distribution of the gravitational potential.   
Recently, Madsen et al. (1997) have demonstrated by N-body  
simulations that the under dense and the over dense regions are 
closely linked to the regions with the positive and the negative 
gravitational potential respectively.
Thus, given all these results showing the important role of 
the primordial gravitational potential in the structure formation,
it would be interesting to calculate the effect of the primordial 
potential upon the mass distribution function. 

The mass distribution 
function $n(M)$ is defined such that $n(M)dM$ is the comoving 
number density of gravitationally bound objects 
in the mass range $(M,M + dM)$.
The standard Press-Schechter (hereafter, PS) formalism 
provides an effective tool to evaluate $n(M)$ 
in spite of various criticism on it (see \cite{mon98}),
and is widely used in cosmology 
(e.g. \cite{gro-etal97}; \cite{kit-sut97}; \cite{bah-fan98}; 
\cite{rob-gaw-sil98}; \cite{wan-ste98}).
Also, \cite{lee-sh98} have shown by 
applying the dynamics based on the
Zel'dovich approximation to the PS formalism that it 
is very robust with respect to the underlying dynamics.

The following 
two equations represent the essence of the PS formalism
(\cite{pre-sch74}): 
\begin{equation} 
n(M) = \frac{\bar{\rho}}{M}\bigg{|}
\frac{dF}{dM}\bigg{|} ,
\end{equation}
\begin{equation}
F(M) = \int^{\infty}_{\delta_{c}}\! p(\delta) d\delta.  
\end{equation}
Here $p(\delta)$ is the probability density distribution of the 
linearly extrapolated density contrast $\delta$ 
smoothed on a comoving filtering 
scale $R$ which is related to the mass by 
$M=M(R)=\alpha{\bar\rho}R^3$. 
The proportionality constant $\alpha$ 
is either determined by the shape of
the smoothing window function or sometimes 
is used as a free parameter in order to provide 
a better agreement with numerical results. 
In the case of a sharp k-space filter which is actually 
consistent with the PS formalism (see \cite{pea-hea90}), 
the filtering scale $k_c = 2\pi/R$ in k-space 
and mass are
related as $M =  6\pi^{2}\bar{\rho}k_c^{-3}$.
The density threshold value $\delta_c$ for collapse 
was originally given as $\delta_{c}\approx 1.69$    
according to the Top Hat spherical model.  
However, it has been shown that the lowered value of 
$\delta_c$ in the range from $1.3-1.6$ 
gives a better fit in N-body simulations, which depends on 
the the initial spectrum and the type of the filter 
(e.g.,  \cite{gro-etal97}). 

In this Letter we investigate and show how much the primordial 
gravitational potential $\varphi$ 
affects the mass distribution function of galaxy cluster.   
Modifying the PS formalism, 
we derive  a constrained mass distribution
function $n(M|\varphi$) defined 
as the comoving number densities of 
clumps of mass $M$ in the regions where the primordial gravitational
potential fluctuation satisfies some specified conditions.  
The Cold Dark Matter model (CDM) 
with $\Gamma = \Omega  h = 0.25$ and $\sigma_8=1$  
is used to demonstrate the significance of the effect.

\section{MODIFICATION OF THE PS FORMALISM}

In order to incorporate the 
primordial gravitational potential fluctuations
term into the above equations, 
we first derive the conditional probability density distribution 
$p(\delta|\varphi<-\varphi_{c})$ ($\varphi_{c}>0$):
\begin{eqnarray}
p(\delta|\varphi<-\varphi_{c}) &=& 
\frac{1}{\sqrt{2\pi}\sigma_{\delta}}
\exp\Bigg{(}-\frac{\delta^2}{2\sigma_{\delta}^2}\Bigg{)}
\Bigg{[}1-{\rm erf}\bigg{(}
\frac{\varphi_{c}}{\sqrt{2}\sigma_{\varphi}}
\bigg{)}\Bigg{]}^{-1} \times \nonumber \\ 
&&\Bigg{[}1+{\rm erf}\Bigg{(}\frac{\kappa
\frac{\delta}{\sigma_{\delta}}- \frac{\varphi_c}{\sigma_{\varphi}}}
{\sqrt{2(1-\kappa^2)}}
\Bigg{)}\Bigg{]}.
\end{eqnarray}
Here $\sigma_{\delta}^2$, $\sigma_{v}^2$, and $\sigma_{\varphi}^2$
are the density, velocity and the potential 
variances respectively; $\kappa = <\delta\cdot\varphi>/ 
\sigma_{\delta}\sigma_{\varphi} =
\sigma_{v}^2/\sigma_{\delta}\sigma_{\varphi}$ is 
the crosscorrelation coefficient of the the density contrast 
$\delta$ smoothed on the scale $k_c$ and the primordial 
({\it unsmoothed}) potential fluctuations $\varphi$.
The probability density
distribution function for 
$\varphi > \varphi_c$ has the same 
form as equation (3) except for the opposite sign 
in front of the second error function term. 
 
In the case of a sharp k-space filter
assumed throughout this Letter, 
$\sigma_{\delta}^2$, $\sigma_{v}^2$ and $\sigma_{\varphi}^2$  
are given as  
\begin{eqnarray}
&&\sigma^{2}_{\delta}(M) = \frac{1}{2\pi^2}
\int^{k_{c}(M)}_{0}\! dk k^{2}P(k), \\
&&\sigma^{2}_{v}(M) = \frac{1}{2\pi^2}
\int^{k_{c}(M)}_{0}\! dk P(k), \\
&&\sigma^{2}_{\varphi} = \frac{1}{2\pi^2}
\int^{\infty}_{k_{l}}\! dk k^{-2}P(k), 
\end{eqnarray}
where $P(k)$ is the density power spectrum. 
It is worth stressing that 
the variances $\sigma^{2}_{\delta}$ and $\sigma^{2}_{v}$ depend on $k_c$,
but the variance of the primordial potential, 
$\sigma_{\varphi}^2$ does not. 
We are interested in evaluating
the effect of the {\it primordial} potential fluctuations and thus
take $\sigma_{\varphi}$  with no filtering.  
The long wave cutoff $k_{l} \approx
3000 h^{-1} {\rm Mpc}$ (eq. 6)
corresponds to the assumption that the waves longer
than the cosmological horizon are irrelevant to the processes 
on scales of the structures in the universe (i.e. smaller than 
a few hundred of $h^{-1} {\rm Mpc}$). 
A particular choice of $k_{l}$ is unimportant since 
for the Harrison-Zel'dovich spectrum (assumed here)
$\sigma^{2}_{\varphi}$ depends on $1/k_{l}$ only logarithmically.

As a result of the incorporation of  the potential into the
probability density distribution,  the volume fraction 
$F(M|\varphi<-\varphi_{c})$   
is now a function of both
$\sigma_{\delta}$ and $\sigma_{v}$ 
each of which in turn is a function of mass M:
\begin{equation}
F(M|\varphi<-\varphi_c) = F[\sigma_{\delta}(M), \sigma_{v}(M)]  
= \int^{\infty}_{\delta_{c}}\! 
d\delta ~p(\delta|\varphi < -\varphi_{c}).
\end{equation}
Thus, equation (1) for the conditional mass function 
$n(M|\varphi<-\varphi_{c})$ becomes
\begin{equation}
n(M|\varphi<-\varphi_{c}) = \frac{\bar{\rho}}{M}
\Bigg{|}\frac{\partial F}{\partial\sigma_{\delta}}
\frac{d\sigma_{\delta}}{dM} 
+ \frac{\partial F}{\partial\sigma_{v}}
\frac{d\sigma_{v}}{dM}\Bigg{|}.
\end{equation}

For the initial power spectrum $P(k)$, we use the fit 
given by \cite{bar-etal86}  
with $\Gamma = \Omega h = 0.25$ and $\sigma_8= 1$ normalization.  
This choice of the parameters is in general agreement with both 
the COBE measurements and the galaxy two-point correlation
function (\cite{pea-dod94}). 
Calculating $d\sigma_{\delta, v}/dM$ numerically, 
we evaluate $n(M|\varphi<-\varphi_{c})$ through equations (4) to (8). 
We reserve the detailed description of the calculation 
for a companion paper in preparation. 

Fig. 1  shows the magnitude of the effect in terms of the cumulative
mass functions. The upper panel 
shows the cumulative conditional mass functions 
$n(>M|\varphi<-\varphi_{c})$ and $n(>M|\varphi>\varphi_{c})$ 
(for each case of $\varphi_{c}=\sigma_{\varphi}, 0$) along with
the unconditional PS mass function,  $n(>M)$.
The cumulative mass functions were obtained by integrating 
the  mass functions $n(M|\varphi<-\varphi_{c})$ numerically. 
The shaded area shows the fit to the observational mass function
of rich clusters given by \cite{bah-cen93}.
The ratios of the conditional 
mass functions to the unconditional one are plotted
in the lower panel of Fig. 1.  
 
We also calculate the probability that a clump
with mass M is located in the potential regions satisfying the 
chosen condition, for instance,  $\varphi<-\varphi_{c}$  
\begin{equation}
P(\varphi<-\varphi_{c}|M) = \frac{n(M|\varphi<-\varphi_{c})}
{n(M)}P(\varphi<-\varphi_{c}),
\end{equation}
where $P(\varphi<-\varphi_{c})$ is the fraction of space
satisfying the given condition (see Fig. 2). 
The other probabilities $P(-\varphi_{c}<\varphi<0|M)$, 
$P(0<\varphi<\varphi_{c}|M)$, and $P(\varphi_{c}<\varphi|M)$
can be obtained in a similar manner. 
   
\section{RESULTS AND DISCUSSION}
The PS formalism has been proved to be a simple but very effective tool 
widely used for constraining cosmological models. 
We have modified it by considering 
the dependence of mass function on 
the initial perturbation of gravitational potential.  
The resulting modified PS theory predicts that the clumps with 
masses greater than roughly $10^{14}h^{-1}M_{\odot}$ 
have a noticeable tendency to form in the troughs of the
primordial gravitational potential (the regions where the
primordial potential fluctuations were negative). 
This quantitative prediction 
can be tested in large N-body simulations. 
Regardless of the outcome
it will shed light on the PS formalism;
if our prediction is confirmed, it will show a new potency of the
PS technique. Otherwise a new limitation to the formalism 
will be established.
  
Assuming that the prediction is correct at least qualitatively, 
\footnote{N-body simulations
(e.g., Madsen et al. 1997) 
and the adhesion model (Sahni et al. 1994) 
have already visually demonstrated this bias effect of 
the gravitational potential.}
we would like to discuss some of its obvious consequences.
The scale of the initial potential 
\begin{equation}
R_{\varphi}
= \sqrt{3} \sigma_{\varphi}/
\sigma_{\varphi'} 
=\sqrt{3\frac{\int^{\infty}_{k_{l}}\! dk k^{-2}P(k)}
{\int^{\infty}_{0}\! dk P(k)}}
\approx 120 h^{-1} {\rm Mpc}
\end{equation}
does not depend on any ad hoc scale; the dependence on $k_l$ is
exremely weak ($\propto \sqrt{ln{(1/k_l)}}$ for the Harrison-Zel'dovich
spectra assumed here).
It is, perhaps,
worth mentioning that the scale of the potential is also practically 
independent of the smoothing scale unless it exceeds 
the value of a few tens of $h^{-1}{\rm Mpc}$.
The density scale $R_{\delta_{k_c}}$ is determined by the scale 
of the smoothing window function $k_c$ that has only one ``natural''
scale corresponding nonlinearity $k_c=k_{nl}$.
For the model in question the scale of the primordial potential
is found to be $R_{\varphi} \approx 120 h^{-1} {\rm Mpc}$. 
The scale of the density contrast field reaches this value 
$R_{\delta} = \sqrt{3} \sigma_{\delta}/\sigma_{\delta'} 
\approx 120 h^{-1} {\rm Mpc}$ 
only after it is smoothed  on $k_c \approx 0.017 h {\rm Mpc^{-1}}$.  
The corresponding density variance on this scale is 
$\sigma_{\delta}(0.017 h {\rm Mpc^{-1}}) \approx 0.03$.
On the other hand, the number of clumps
with masses $10^{14} - 10^{15} h^{-1} M_{\odot}$ can
easily be  30\% greater in the troughs of the potential than 
the mean density 
$n(>M) = 0.5[n(>M|\varphi<0)+ n(>M|\varphi>0)]$ 
(see Fig. 1). Thus, the bias factor $b$ (defined by the relation
$\Delta n_{cl}/ n_{cl} = b \Delta \rho_m/\rho_m$) 
reaches at least $10$ on the scale about $120 h^{-1} {\rm Mpc}$.

Qualitatively the bias phenomenon  can be explained as follows. 
The initial density
contrast is proportional to the Laplacian of the initial potential
($\delta \propto  \nabla^2 \varphi$). 
Therefore the two fields are cross-correlated:  
the positive peaks of $\delta$ are more likely 
to be found in the troughs of the potential where it is negative. 
The correlation is not very strong 
(for $k_c = 0.25 h {\rm Mpc^{-1}}$ corresponding
to $\sigma_{\delta} = 1$ the crosscorrelation coefficient
$\kappa = \sigma_v^2(0.25 h {\rm Mpc^{-1}})/ 
\sigma_{\varphi} \sigma_{\delta}(0.25 h {\rm Mpc^{-1}})
\approx 0.12$).  But the clusters are extreme
objects corresponding to the tail of the mass function, 
and thus very sensitive 
to the environment. That is why the clusters put one of the strongest 
constraints on cosmological models 
(\cite{kly-rhe94}, \cite{bo-my96}, \cite{fan-etal97}, \cite{bah-fan98}).  

Incorporating the motion of mass into dynamics can only increase
the bias effect due to the nonlinear effects although they are quite small
on the scale in question. But, the point is not in the magnitude
of the nonlinear effects but rather in their sign.
On the scale of the potential
the mass moves from the peaks of the potential to the troughs. 
Using the Zel'dovich approximation one can easily estimate 
the rms displacement of the mass  on the scale of the potential
(\cite{sh93}): 
\begin{equation}
d_{rms} = \sqrt{{{\int_0^{0.017h}P(k) dk} \over {\int_0^{0.25h}P(k)k^2 dk}}}
\approx 3 h^{-1} {\rm Mpc}. 
\end{equation}
It is relatively small compared to the scale of the potential but 
coherent on the scale of the potential field, 
and therefore it can only enhance the bias effect.
Another nonlinear effect is related to the rate of growth of 
perturbations.
For the perturbations on the scale of a few Mpc the potential 
troughs/peaks may be viewed as patches with slower/faster
expansion rate that corresponds to the increase/decrease
of the rate of growth of small-scale perturbations.
Similarly, the bias is enhanced in the redshift space
because the velocity field is directed toward the 
troughs and away from peaks of the potential. Both effects can
increase the bias by about 5\% depending on the initial spectrum.

Another  way of calculating the constrained mass 
function would be using the peak-background split technique  
suggested by \cite{kai84} to explain the enhanced correlation 
function of reach clusters. 
Obviously, the initial potential resembles the smoothed initial 
density field if the filter has a sufficiently large scale,
but the former is never identical to the latter. 
The potential itself can be viewed as
a smoothed density field with a very soft scale-free filter 
$W(k) \propto k^{-2}$. Typically the density field 
is filtered with much harder filters (e.g. top-hat, Gaussian, 
or sharp $k$-space filters), that impose the scale which is an ad hoc
parameter.
The magnitude of the bias in our approach is determined by the 
crosscorrelation of the density contrast smoothed at the scale ($k_c$) of 
nonlinearity ($\sigma_{\delta_{k_c}}=1$) with the initial potential
that does not have any ad hoc parameters.
Probably, the value of the crosscorrelation coefficient
determines the bias in the peak-split approach as well.
The crosscorrelation of the density field $\delta_{k_c}$ 
smoothed on the scale of nonlinearity ($k_c = 0.25 h {\rm Mpc^{-1}}$)
with the field $\delta_{k_{\varphi}}$ smoothed on the scale 
of the potential ($k_{\varphi} = 0.017 h {\rm Mpc^{-1}}$) is
about $4$ times weaker than the correlation of $\delta_{k_c}$ 
with the initial potential $\varphi$.
Thus, we expect that the bias of galaxy clusters
on such large scales as the scale of the initial potential ($\approx
120 h^{-1} {\rm Mpc}$) is stronger toward the troughs of the 
potential fluctuations rather than to the peaks of 
the density fluctuations $\delta_{k_c}$ smoothed
with the corresponding filter. We have not applied the split
peak-background approach because it is not clear 
how to avoid arbitrarines in choosing the scale that splits
the density into small-scale peaks and large-scale
background field. This question requires a separate study.  

Applying this effect to observations one has to
take into account the following issues. 
The gravitational potential does not evolve much on large scales
especially in the Einstein-de Sitter universe 
(Kofman \& Shandarin 1988; \cite{pau-mel95}; \cite{mel-etal96}).  
Therefore, the potential
at present is very similar to the primordial one on scales much greater
than the scale of nonlinearity. 
A simple explanation to this in the frame of the standard scenario
of the structure formation is due to the fact that the mass 
has been displaced by the distance about $10 h^{-1}{\rm Mpc}$
(\cite{sh93}). Therefore, the potential on  scales
greater than, say, $30 h^{-1}{\rm Mpc}$ has been changed very little.

Clusters can be used as {\it statistical} tracers of the potential. 
In addressing this question it is worth noting that  
the shot noise is an important factor since clusters are rare objects. 
Using the observational mass function (Bahcall \& Cen 1993)
one can estimate that an average spherical patch  
of the radius  $\approx 60 h^{-1} {\rm Mpc}$
contains about $30$ clusters with the
masses greater than $10^{14} h^{-1} M_{\odot}$. Thus, the shot noise 
is about 18\% on this scale which is comparable with the bias
itself (see Fig. 1, the bottom panel). However, Fig. 2 suggests that
the most massive clusters [$M>10^{15} h^{-1} M_{\odot}$] are very
likely to reside in the regions of negative potential ($P > 75\%$)
and very unlikely in the regions of high potential ($P < 5\%$
if $\varphi > \sigma_{\varphi}$). More detailed analysis will be present
elsewhere. 

Probably, the best candidates for the markers of the troughs in the
field of the primordial potential fluctuations 
are superclusters (defined as clusters of clusters)
(\cite{bah-son84}) especially with highest density enhancements 
(Shapley supercluster) and the giant
geometrical patterns in the cluster distribution (\cite{tul-etal92}).

\acknowledgments

We wish to thank Anatoly Klypin for 
useful comments and helpful discussions. 
We acknowledge the support of NASA grant NAG 5-4039 and EPSCoR 1998 grant.

\begin{figure}[tb]
\psfig{figure=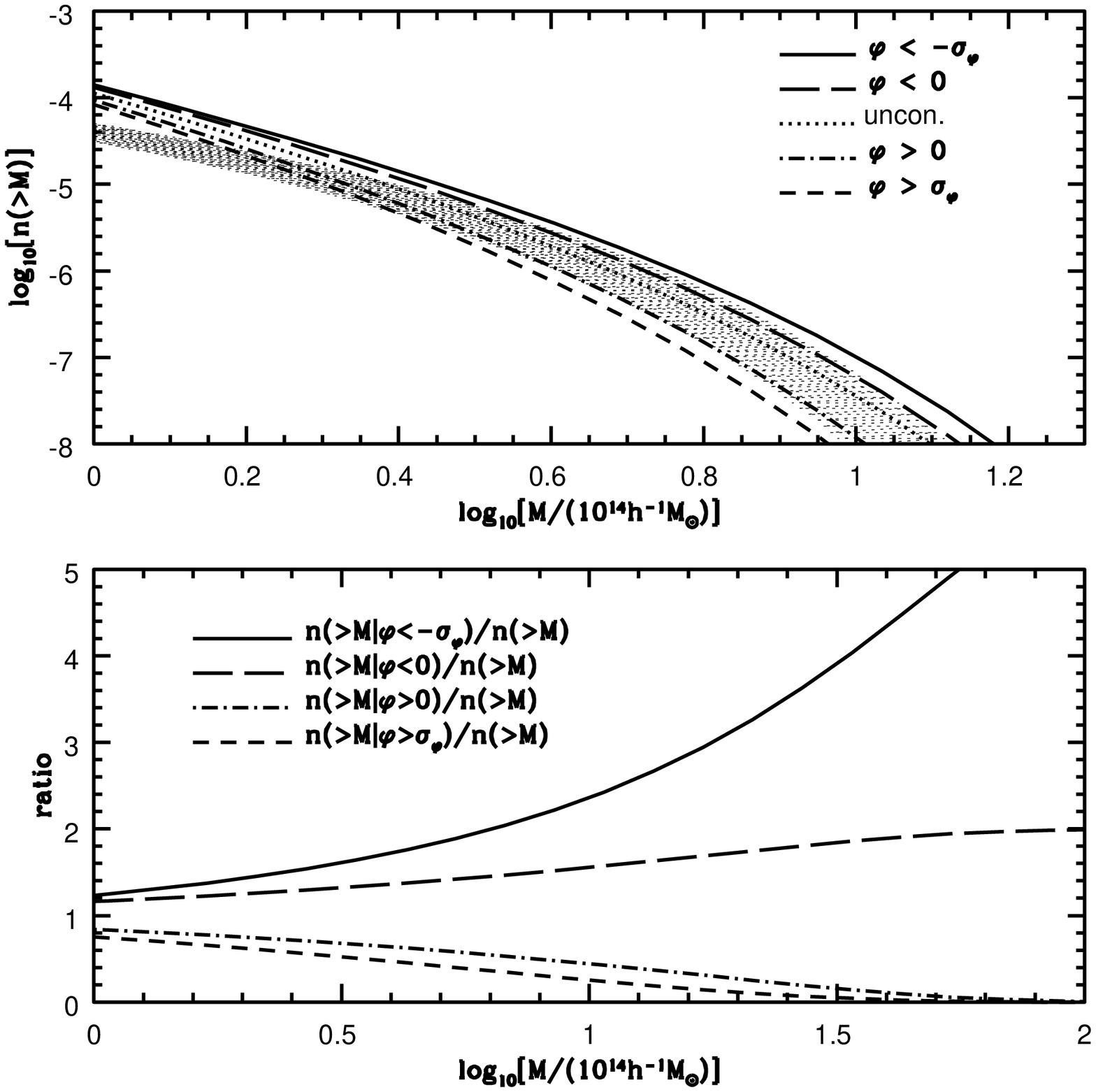,height=14.5cm,width=15.5cm}
\caption[fig1.ps]{In the upper panel the conditional cumulative  
mass function satisfying chosen potential condition is plotted. 
The solid, the long dashed, the 
dot-dashed, and the dashed lines correspond to the conditions  
$\varphi  < -\sigma_{\varphi}$, $\varphi  < 0$, 
$\varphi  > 0$, and  $\varphi  > \sigma_{\varphi}$ respectively, 
while the dotted line represents the unconditional cumulative 
PS mass function.  The shaded area is $1\sigma$ fit to 
the observational cumulative mass function of rich clusters
by Bahcall and Cen (1993).
In the lower panel the ratio of the 
conditional  cumulative mass functions to the unconditional 
one is plotted for each condition. 
The CDM spectrum with $\Gamma = 0.25$ normalized to 
$\sigma_{8} = 1$ has been used.}
\end{figure}
\begin{figure}[tb]
\psfig{figure=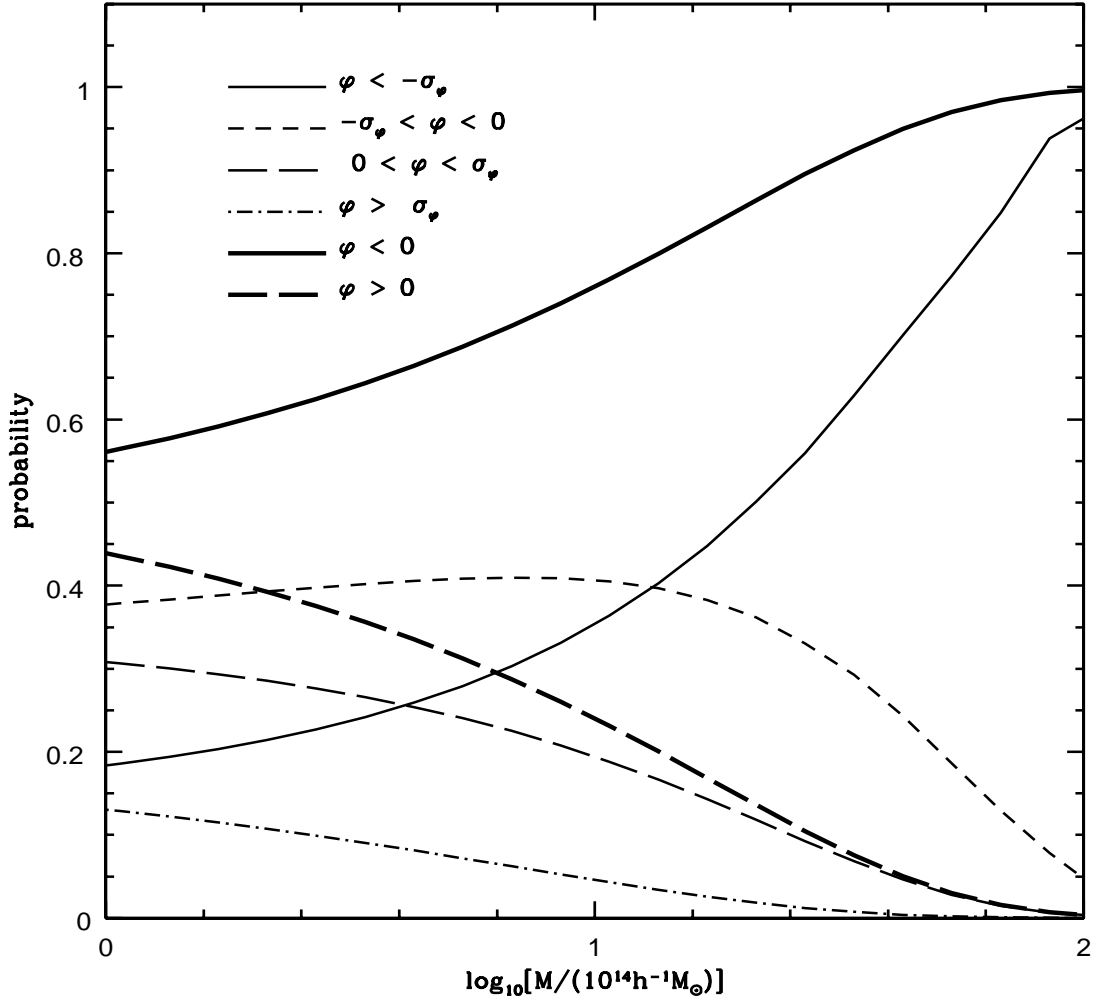,height=14.5cm,width=15.5cm}
\caption[fig2.ps]{The probability that a
clump with mass M can be found in the regions satisfying chosen
potential condition is plotted. The heavy solid, the 
heavy dashed, the solid, the dashed, the 
long dashed, and the dot-dashed lines correspond to the 
condition $\varphi  < 0$, $\varphi  > 0$,
$\varphi  < -\sigma_{\varphi}$, $-\sigma_{\varphi} < \varphi  < 0$, 
$0 < \varphi < \sigma_{\varphi}$, and   
$\varphi  > \sigma_{\varphi}$ respectively. }
\end{figure} 
\end{document}